# The effect of Fe and Ni catalysts on the growth of multiwalled carbon nanotubes using chemical vapor deposition

Joydip Sengupta · Chacko Jacob

**Abstract** The effect of Fe and Ni catalysts on the synthesis of carbon nanotubes (CNTs) using atmospheric pressure chemical vapor deposition (APCVD) was investigated. Field emission scanning electron microscopy (FESEM) analysis suggests that the samples grow through a tip growth mechanism. High-resolution transmission electron microscopy (HRTEM) measurements show multiwalled carbon nanotubes (MWCNTs) with bamboo structure for Ni catalyst while iron filled straight tubes were obtained with the Fe catalyst. The X-ray diffraction (XRD) pattern indicates that nanotubes are graphitic in nature and there is no trace of carbide phases in both the cases. Low frequency Raman analysis of the bamboo-like and filled CNTs confirms the presence of radial breathing modes (RBM). The degree of graphitization of CNTs synthesized from Fe catalyst is higher than that from Ni catalyst as demonstrated by the high frequency Raman analysis. Simple models for the growth of bamboo-like and tubular catalyst filled nanotubes are proposed.

## Introduction

Carbon nanotubes (CNTs) provide significant advantages over numerous existing materials due to their unique mechanical, electronic, thermal, and chemical properties (Chen et al. 1999; Dai et al. 1996; Qian et al. 2000; Saito et al. 1997; Tans et al. 1998) which have made CNTs a center of attraction in the field of nanoscale research since their discovery in 1991 by Iijima (1991). The unprecedented interest in a variety of possible commercial applications has accelerated research and development in the area of CNTs. The synthesis of CNTs can be grouped into the following categories (Iijima 1993; Journet et al. 1997; Kong et al. 1998; Meyyappan et al. 2003; Thess et al. 1996): arc discharge, chemical vapor deposition, plasma method, laser ablation, etc. Among the above methods, CVD method is simple and easy to implement, and has been widely used because of its potential advantage to produce a large amount of CNT growing directly on a desired substrate with high purity, large yield and controlled alignment, whereas the nanotubes must be collected separately in the other growth techniques.

The general catalytic growth process of CNTs by CVD is based on the following mechanism proposed by Baker et al. (1972) known as the vapor–liquid–solid (VLS) mechanism. In this mechanism, the liquid catalytic particles at high temperature absorb carbon atoms from the vapor to form a metal-carbon solid state solution. When this solution becomes supersaturated, carbon precipitates at the surface of the particle in its stable form and this leads to the formation of a carbon tube structure. The metal clusters which act as a catalyst for CNT growth by CVD can be produced by different methods, e.g., by vapor or sputter deposition of a thin metallic layer on the substrate, by saturation of porous materials by metals, by deposition of solutions of substances containing catalyzing metals, or by introduction of organometallic substances into the reactor (Bertoni et al. 2004; Huang et al. 2003; Jeong et al. 2002; Tu et al. 2002). Prior to CNT synthesis, high temperature hydrogen treatment is an important step in order to produce contamination free catalyst particles and for the removal of oxides that may exist over the catalyst surface (Takagi et al. 2007).

In the catalytic method, nickel and iron are widely used as pure-metal catalysts for CNT growth. So a detailed comparative study on the effects of different metal catalysts on CNT growth, morphology, and microstructure is of importance. Despite tremendous progress in synthesizing CNTs, the systematic comparative study of the catalyst effect on the CNT growth is still not much reported yet. Exact understanding of the catalyst activity would eventually lead to a controlled growth of CNTs, which is a prerequisite for various potential applications such as bamboo-like CNTs are good electric conductive additive for lithium ion batteries (Zou et al. 2008) while for magnetic recording magnetic particle-embedded CNTs are preferable (Kuo et al. 2003).

Here, we report a systematic study of the effect of nickel and iron catalysts on the synthesis of CNTs by APCVD over Si (111) under identical conditions, using propane as a source of carbon. Propane was used as a readily available high purity hydrocarbon source for CNT growth. The catalyst distribution was studied by AFM (Atomic force microscopy). The morphology, internal structure, and degree of graphitisation of CNTs grown on Fe and Ni catalyst particles were investigated using FESEM, HRTEM, XRD, and Raman spectroscopy. To the best of our knowledge, the low frequency Raman analysis of the bamboo-like and filled CNTs is reported for the first time which confirms the presence of RBM. We observed that the internal structure and degree of graphitization of CNTs depend on the catalyst species, providing a method to modify CNT growth according to the catalyst.

# Experimental details

APCVD of CNTs was carried out by catalytic decomposition of propane on Si (111) wafers with a pre-treated catalyst overlayer in a hot-wall horizontal reactor using a resistance-heated furnace (ELECTROHEAT EN345T). The Si (111) substrates were ultrasonically cleaned with acetone and deionised water prior to catalyst film deposition. For the deposition of the catalyst film, the respective metals were loaded in a vacuum system (Hind Hivac: Model 12A4D) and pumped down to a base pressure of $10^{-5}$ torr and catalyst films ($\sim 20$ nm thick) were deposited by evaporation.

The substrates were then loaded into a quartz tube furnace, pumped down to $10^{-2}$ torr and backfilled with flowing argon to atmospheric pressure. The samples were then heated in argon up to 900 °C following which the argon was replaced with hydrogen. Subsequently, the samples were annealed in hydrogen atmosphere for 10 min. Finally, the reactor temperature was brought down to 850 °C and the hydrogen was turned off, thereafter propane was introduced into the gas stream at a flow rate of 200 SCCM, for 1 h for CNT synthesis. The CNTs synthesized using the Fe catalyst were assigned the name F-CNT and the CNTs synthesized with Ni catalyst were assigned the name N-CNT.

An AFM (Nanonics Multiview 1000$^{TM}$) system was used to image the surface morphology of the catalyst layer before and after the heat treatment with a quartz optical fiber tip in intermittent contact mode. Samples were also characterized by a Philips X-Ray diffractometer (PW1729) with Cu source and $\theta$–2$\theta$ geometry to analyze the crystallinity and phases of grown species. Micro Raman measurements were carried out at room temperature in a backscattering geometry using a 488 nm air-cooled Ar$^+$ laser as an excitation source for compositional analysis. We have used an assembled Raman spectrometer; the spectrometer was equipped with a TRIAX550 single monochromator with a 1,200 grooves/mm holographic grating, a holographic super-notch filter and a Peltier-cooled CCD detector. FESEM (ZEISS SUPRA 40) and HRTEM (JEOL JEM 2100) equipped with an EDX analyzer (OXFORD instrument) were employed for examination of the morphology and microstructure of the CNTs.

## Results and discussion

Figure 1a, b shows the 3-D AFM images of the distribution of the as-deposited Fe and Ni catalyst particles over the Si (111) substrate. The AFM image reveals that in both cases the initial films consist of clusters instead of a continuous layer. We observed that the deposition process itself creates a relatively rough surface. As the catalyst atoms interact more strongly with themselves than they do with the Si substrate in both the cases, hence the catalyst deposition proceeds via island nucleation and coalescence (Tiller 1991).

After deposition of the catalyst, the samples were annealed in hydrogen for 10 min at 900 °C. This heat treatment results in the formation of islands as confirmed by AFM images in Fig 1c, d for Fe and Ni catalyst film, respectively. Heating above a certain temperature causes clusters to coalesce and form macroscopic islands. This process is based on cluster diffusion and depends on their density and diffusion coefficient, at a given substrate temperature. Cluster diffusion terminates when the island shape is of minimum energy for the specific annealing conditions (Jak et al. 2000). These clusters act as a catalyst surface for nanotube growth.

FESEM was employed for the analysis of the morphology and density of CNTs. The FESEM micrographs in Fig. 2a, b shows the surface morphology of F-CNT and N-CNT, respectively. High-aspect ratio nanostructures are observed on the two catalyst film surfaces. The area density of the deposited CNTs was high and the CNT structures had randomly oriented spaghetti-like morphology in both cases. In many cases, small bright catalyst particles were detected at the tip of the F-CNTs and N-CNTs (right insets of Fig. 2a, b). This suggests that the tip growth mechanism is likely to be responsible for the CNT synthesis under the present conditions for both the catalysts. The EDX spectrum obtained from F-CNT and N-CNT samples are shown as left insets in the Fig. 2a, b. The atomic ratio of C:Fe is ∼175 and the atomic ratio of C:Ni is ∼422, as derived from the EDX analysis.

HRTEM was used to characterize the growth morphology and structure dependence of the nanotubes on the catalysts. The sample preparation for HRTEM study was done by scraping the nanotubes from the Si substrates and dispersing them ultrasonically in alcohol and transferring to carbon coated copper grids. The high-resolution transmission electron microscope (HRTEM) images of F-CNT and N-CNT, respectively are shown in Fig. 3a, b. For F-CNT with diameter of about 20 nm, a few elongated particles can be observed embedded into the core of the nanotube (Fig. 3a), which demonstrates the hollow nature of the nanotube deposited using Fe. The average size of these particles is around 15 nm. Chemical composition analysis (EDX) confirms (not shown here) that the elongated particles are of iron. A bundle of such nanotubes can be seen in the upper inset of Fig. 3a with outer diameter distribution around 10 nm to 30 nm and the lower inset image shows the arrangement of graphitic walls of F-CNT.

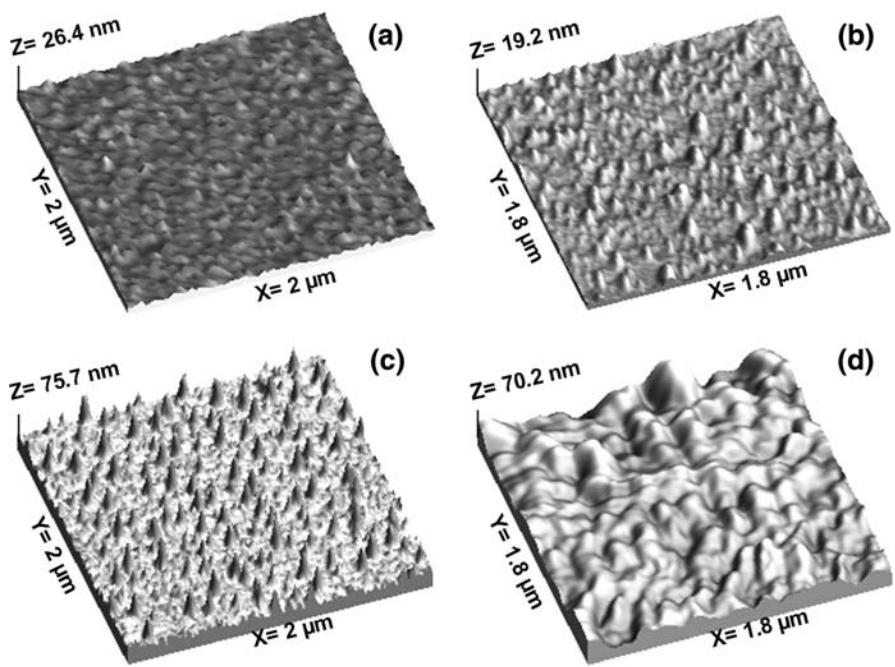

**Fig 1** **a** A 3D AFM image of the as-deposited Fe film, showing the clusters forming at the early stages of growth, **b** A 3D AFM image of the as-deposited Ni film, showing the clusters forming at the early stages of growth, **c** A 3D AFM image of Fe islands formed upon annealing the substrate at 900 °C, **d** A 3D AFM image of Ni islands formed upon annealing the substrate at 900 °C

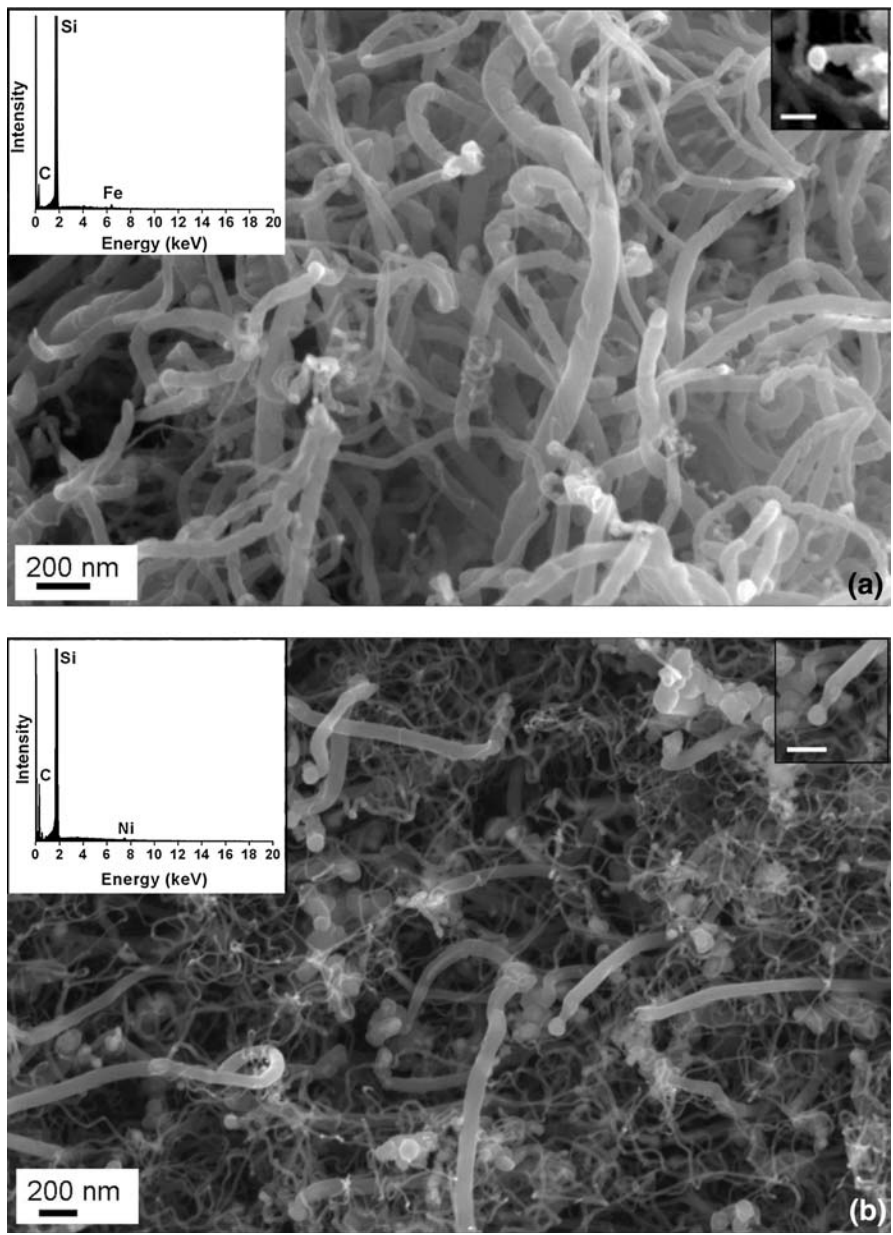

Fig 2 **a** FESEM micrograph of the as-grown MWNTs deposited by the APCVD method using Fe catalyst, (*left inset*) the EDX spectrum obtained from F-CNT, (*right inset*) small bright catalyst particles were detected at the tip of the F-CNT and the scale bar length is 200 nm. **b** FESEM micrograph of the as-grown MWNTs deposited by the APCVD method using Ni catalyst, (*left inset*) the EDX spectrum obtained from N-CNT, (*right inset*) small bright catalyst particles were detected at the tip of the N-CNT and the scale bar length is 200 nm

The upper inset of Fig. 3b shows the typical HRTEM image of the N-CNTs and the lower inset image shows the arrangement of graphitic walls of N-CNT. The results indicate that the carbon nanostructures are bamboo-like MWCNTs with outer diameter distribution around 20 nm to 40 nm. Figure 3b presents the HRTEM image corresponding to an individual nanotube as shown in the upper inset

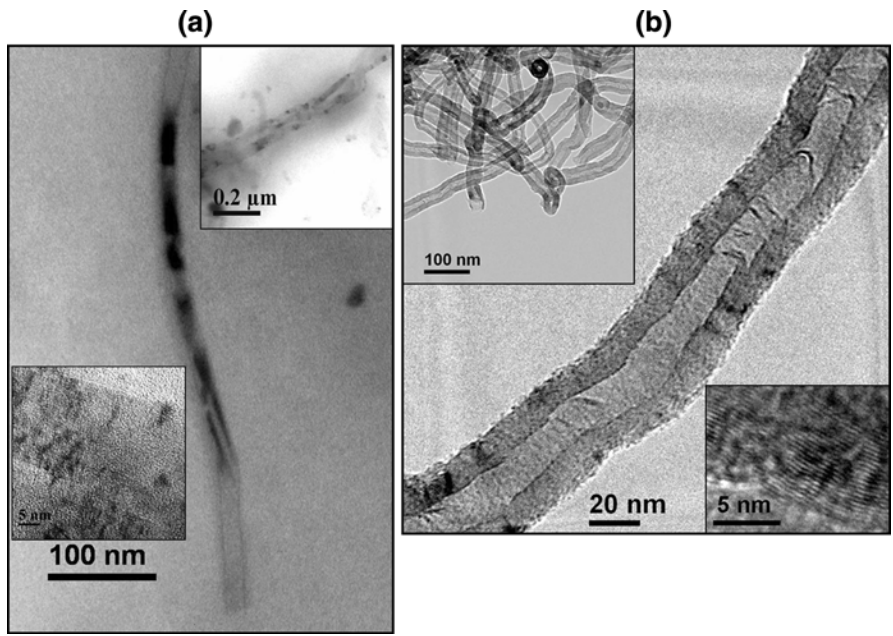

**Fig 3** a HRTEM image of the iron encapsulated CNT grown by the APCVD method using Fe catalyst, (*upper inset*) HRTEM image of the iron encapsulated CNTs grown by the APCVD method using Fe catalyst, (*lower inset*) HRTEM image of interlayer spacing of graphitic carbon in F-CNT. b HRTEM image of the bamboo-like CNT grown by the APCVD method using Ni catalyst, (*upper inset*) HRTEM image of the bamboo-like CNTs grown by the APCVD method using Ni catalyst, (*lower inset*) HRTEM image of interlayer spacing of graphitic carbon in N-CNT of Fig. 3b. The N-CNT is well graphitized with an inner diameter of about 14–17 nm and outer diameter 38–42 nm. The thickness of the tube wall lies in the range of 12–14 nm, which suggests that the tube wall is composed of approximately 30–40 graphitic layers. The compartments of the layers in the bamboo-like structure are clearly shown. The major difference that can be observed from the HRTEM images is that catalyst particles are included within straight nanotube in the case of F-CNTs whereas bamboo-like structures are produced for N-CNTs.

XRD measurements were carried out to examine the structure of the CNTs and the resulting $\theta$–$2\theta$ scan is shown in Fig. 4. The peak at 26.2° is the characteristic graphitic peak arising due to the presence of MWNTs in the sample. For F-CNT, the peak near 43.7° is attributed to the (101) plane of the nanotube and the peak at 44.7° is from the Fe catalyst (JCPDS card No. 06-0696). The intensity of the main CNT peak (002) is lower in case of N-CNT which may be the reason for absence of the (101) peak in its diffraction pattern. The broader FWHM of the (002) peak of N-CNT indicates that it has lower crystallinity than F-CNT. The peak at 28.4°, however, is not from the CNTs and is attributed to (111) plane of the Si substrate.

Raman spectroscopy provides more details of the quality and structure of the materials produced. The

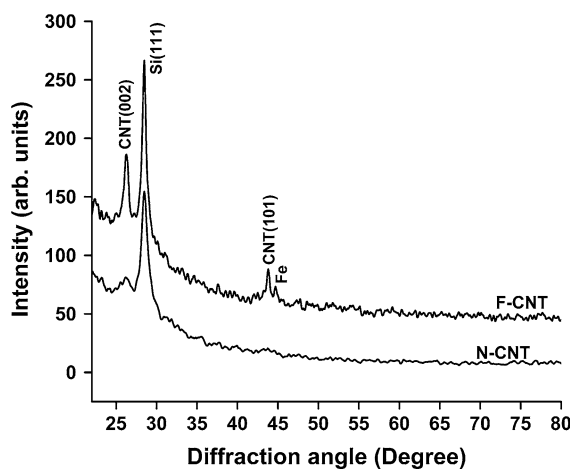

**Fig 4** XRD spectrum of MWNTs grown on Si (111) substrate using Fe and Ni

Raman spectra were taken in the backscattering geometry on the as-grown samples. Figure 5 shows the room temperature Raman spectrum of the MWNT materials at a laser excitation wavelength of 488 nm. The spectrum is divided into two main zones: the low frequency region from 150–400 cm$^{-1}$ (inset of Fig. 5) and the high frequency zone from 1,250–1,750 cm$^{-1}$. The vibrations of CNT originate from the curvature induced strain due to misalignment of the $\pi$-orbitals of adjacent coupled carbon atoms. These vibrations are reflected in the Raman peaks.

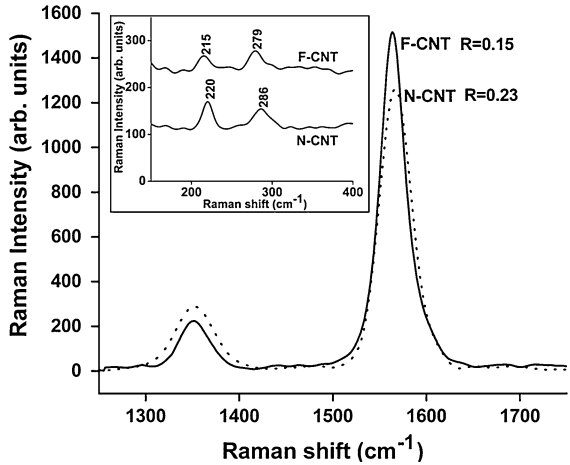

**Fig 5** High frequency Raman spectrum (488 nm excitation) of MWNT films grown by APCVD on Si using Fe and Ni, (Inset) Low frequency Raman spectrum (488 nm excitation) of MWNT films grown by APCVD on Si using Fe and Ni showing the presence of RBM

Evidence for the presence of RBM vibrations in F-CNT and N-CNT samples are obtained from the low wave number range of the spectrum (inset of Fig. 5). This mode has $A_{1g}$ symmetry and all the carbon atoms move in phase in the radial direction creating breathing like vibration of the entire tube (Zhao et al. 2002). The peaks located at 215 cm$^{-1}$ and 279 cm$^{-1}$ in the Raman spectra of F-CNT are due to the RBM of MWNT. The RBM vibrations for N-CNT are located at 220 cm$^{-1}$ and 286 cm$^{-1}$. The frequency of RBM is directly linked to the innermost tube diameter by the relation $\omega_{RBM} = 223.75/d$, where d is the innermost tube diameter and $\omega_{RBM}$ is the wave number in units of nm and cm$^{-1}$, respectively (Jinno et al. 2004). The peaks at 215 cm$^{-1}$, 220 cm$^{-1}$, 279 cm$^{-1}$, and 286 cm$^{-1}$ originate due to the RBM of MWNT with inner-core diameter of 1.04 nm, 1.02 nm, 0.80 nm, and 0.78 nm, respectively. The observation of low frequency modes in the Raman spectra of the as-grown MWNT samples indicates the high purity of the samples (Benoit et al. 2002).

The two main peaks observed in the high frequency zone (Fig. 5) are the so-called D- and G-lines. The G-line corresponds to the $E_{2g}$ mode i.e., the stretching mode of the C–C bond in the graphite plane and demonstrates the presence of crystalline graphitic carbon. For both F-CNT and N-CNT it appears near 1,565 cm$^{-1}$.

The D-line, centered around 1,352 cm$^{-1}$, originates from disorder in the sp$^2$-hybridised carbon and can be due to the presence of lattice defects in the graphite sheet that make up CNTs. The position of D band has been found to vary strongly with the laser excitation energy. The location of the D band for MWNT can be expressed as $\omega_D = 1285 + 26.5 E_{laser}$ (Wei et al. 2003). In our case $E_{laser} = 2.54$ eV resulting in $\omega_D = 1352.3$ cm$^{-1}$, which is in agreement with our observed peak position. However, a recent Raman analysis suggests that the D band is an intrinsic feature of the Raman spectrum of MWNTs, and they are not necessarily an indication of a disordered wall structure (Rao et al. 2000).

In CNT the D/G intensity ratio is reported to increase with increasing structural disorder. Low D/G values are assumed as representative of well-graphitised CNT, thus regarding D/G ratio as a CNT morphology indicator. The values of ($I_D/I_G$) for the CNTs grown using Fe and Ni as catalyst are 0.15 and 0.23, respectively. It indicates that the degree of long-range ordered crystalline perfection of the CNTs grown using Fe is higher than that of CNTs grown using Ni, which is consistent with the XRD results. The increase in the relative intensity of the disordered mode for N-CNT compared to F-CNT samples can be attributed to the increased number of structural defects and sp$^3$-hybridization present in them.

Thus it is observed that the degree of graphitization of CNTs is also dominated by the activity of catalyst.

From the above analysis it can be concluded that for F-CNT and N-CNT, the growth is primarily governed by the tip growth mechanism as proposed by Baker et al. (1972) where, the carbon feedstock molecules are adsorbed on the catalyst particle and are decomposed into carbon atoms that diffuse through the metal particle. Here the diffusion process is most probably driven by a carbon concentration gradient and does not involve the formation of metal carbide (Holstein 1995), in agreement with the XRD and HRTEM studies. From the morphological point of view, the external structure of F-CNT and N-CNT looks quite similar but the main difference lies in their internal structures. The F-CNT has a tubular structure with Fe filling whereas N-CNT has a bamboo-like structure and without metal filling. We suggest the following models as schematically shown in Fig. 6a, b respectively, to explain the growth mechanism of N-CNTs and F-CNTs.

In case of N-CNT, after Ni layer deposition, the catalyst layer becomes fragmented into nanoparticles. The decomposition of propane on the surface of the Ni nanoparticles results in the formation of carbon, and the growth of N-CNTs occurs via diffusion of carbon through the Ni particle. The dissolved carbon diffuses toward the bottom of the Ni particle and the carbon segregates as graphite at the bottom and the side of the Ni, thus forming a hemispherical cap within the nanotube and sealing the existing tube internally. This results in the formation of the knot in the N-CNT, which fully encapsulates the lower part of the catalyst particle. The graphitic layers formed at the bottom surface of the catalyst particle lead to elevation of the Ni particle to the tip. The motive force of pushing out the Ni particle may be caused by the stress accumulated in the graphitic sheath due to the segregation of carbon from the inside of the sheath. If the carbon species are supplied in a steady-state manner, the growth process will repeat and a complete N-CNT with multiple knots along its length will appear (Fig. 6a).

In the majority of reports regarding filled CNT synthesis by CVD (Deck and Vecchio 2005; Kim and Sigmund 2005; Leonhardt et al. 2003; Müller et al. 2006; Zhang et al. 2002), the procedure followed to fill the CNTs is the floating catalyst method, whereas a fixed catalyst film is used for nanotube growth in this article. A plausible mechanism is presented for the F-CNT growth that agrees with the experimental results. The growth process starts with the diffusion of carbon into the metallic iron nanoparticles. Due to the small diameter, the melting point of these nanoparticles is far below the melting point of the bulk metal (Sears and Hudson 1963; Thomas and Walker 1964). This suggests that the nanoparticles are in a liquid state during the decomposition and they can easily change their shape. In particular, the small metal nanoparticles with diameters smaller than the inner tube diameter can diffuse into the cavities due to nanocapillarity (Ajayan and Iijima 1993; Fujita et al. 2003). Constraining forces from the walls of the encapsulating nanotube would confine the catalyst particle to a fixed diameter; this process is shown in Fig. 6b. The diameter restriction provided by the inner nanotube walls also explains why the widths of all the particles observed in the middle of nanotubes matches the inner diameter of the confining CNTs. Segregation of graphite layers takes place when the carbon concentration inside the metallic particle exceeds supersaturation. When the carbon precipitates out, the surface tension of the catalyst increases and the dissolution of carbon starts again leading to CNT growth.

The reason for the different growth morphologies in case of Fe and Ni is not presently clear. However a recent study of the diffusion-controlled growth of MWCNTs revealed that the catalyst particles are in liquid state during tubular growth and in solid state in case of bamboo-like growth (Bartsch et al. 2005). For the capillary action, the presence of liquid-like behavior of the catalyst material at the growth temperature is essential and therefore, the probability of catalyst incorporation

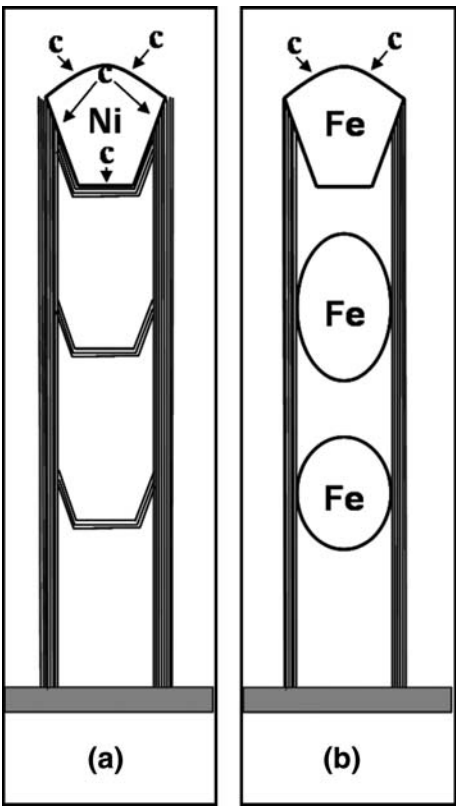

**Fig 6 a** The proposed growth model for the formation of bamboo-like MWNTs using Ni as a catalyst by APCVD, **b** The proposed growth model for the formation of iron-filled MWNTs using Fe as a catalyst by APCVD within the bamboo-like N-CNTs is less than in the tubular F-CNTs.

## Conclusions

The effect of two different catalysts on CNT growth has been studied. The study reveals that the catalyst strongly affects not only crystallinity but also the morphology and microstructure. The CNTs grown on Fe catalyst reveal better degree of graphitization. CNTs growth with Fe and Ni catalysts occurs primarily by the tip growth mechanism but HRTEM studies prove that the internal structure of the grown materials are different, with the N-CNTs being bamboo-like whereas the F-CNTs developing as straight tubes with metal filling. Evidence for the presence of RBM from bamboo-like and metal filled straight tubes was obtained from low frequency Raman analysis. Since the catalyst particles are produced and deposited on a substrate using a dry process, the current method is applicable to any types of substrates. Moreover, many types of nanoparticles can be produced by simple evaporation. This will help in the optimization of catalyst materials for future studies.

**Acknowledgments** The authors are grateful to Dr. A. Roy from the Department of Physics & Meteorology, IIT Kharagpur for her help with the Raman measurement. J. Sengupta is thankful to CSIR for providing the fellowship.